\documentclass[reprint,amsmath,amssymb, aip]{revtex4-2}

\usepackage{float}
\usepackage{comment}
\usepackage{tabularx}

\usepackage[utf8]{inputenc}
\usepackage[T2A,T1]{fontenc}
\usepackage[english]{babel}

\usepackage{color}
\usepackage{ulem}
\usepackage{xcolor}

\usepackage{xfrac}

\usepackage{gensymb}

\usepackage[caption=false]{subfig}

\usepackage{graphicx}
\usepackage{dcolumn}
\usepackage{bm}

\begin{document}

\title{ Nearly perfect routing of chiral light by plasmonic grating on slab waveguide}

\author{Ilia M. Fradkin}
\email{I.Fradkin@skoltech.ru}
\affiliation{Skolkovo Institute of Science and Technology, Bolshoy Boulevard 30, bld. 1, Moscow 121205, Russia}
\affiliation{Moscow Institute of Physics and Technology, Institutskiy pereulok 9, Moscow Region 141701, Russia}
\author{Andrey A. Demenev}
\affiliation{Osipyan Institute of Solid State Physics, Russian Academy of Science, Chernogolovka, 142432, Russia}
\author{Anatoly V. Kovalchuk}
\affiliation{Institute of Microelectronics Technology and High Purity Materials, Russian Academy of Science, Chernogolovka 142432, Russia}
\author{Vladimir D. Kulakovskii}
\affiliation{Osipyan Institute of Solid State Physics, Russian Academy of Science, Chernogolovka, 142432, Russia}
\author{Vladimir N. Antonov}
\affiliation{Skolkovo Institute of Science and Technology, Bolshoy Boulevard 30, bld. 1, Moscow 121205, Russia}
\author{Sergey A. Dyakov}
\author{Nikolay A. Gippius}
\affiliation{Skolkovo Institute of Science and Technology, Bolshoy Boulevard 30, bld. 1, Moscow 121205, Russia}

\date{\today}

\begin{abstract}
Grating couplers are widely used to couple waveguide modes with the far field. Their usefulness is determined not only by energy efficiency but also by additional supported functionality. In this paper, we demonstrate a plasmonic grating on a silicon nitride slab waveguide that couples both TE and TM waveguide modes with circularly polarized light in the far field. Specifically, we experimentally confirmed that circularly polarized light excites TE and TM modes propagating in opposite directions, and the direction is controlled by the handedness. The routing efficiency for normally incident light reaches up to 95\%. The same structure operates in the outcoupling regime as well, demonstrating up to 97\% degree of circular polarization, where the handedness is determined by the polarization and propagation direction of outcoupled modes. Our results pave the way for the realization of polarization-division multiplexers and demultiplexers, integrated circular polarization emitters, as well as detectors of the polarization state of the incident optical field.
\end{abstract}

\maketitle

\section{Introduction}

A recent trend in the development of modern photonic devices is the combining of the lump components into the integrated circuit. One of the imminent tasks is an integration of spectrometers and light sources with input and output channels. The focus in this field is primarily on optimizing the coupling efficiency to minimize energy losses~\cite{zaoui2012cost,zaoui2014bridging,taillaert2004compact,michaels2018inverse,sacher2014wide,hong2019high,ding2014fully}. Many studies have been conducted on the application of apodized (slightly aperiodic) gratings~\cite{mekis2010grating,halir2009waveguide,mehta2017precise,taillaert2004compact,chen2010apodized,marchetti2017high,sacher2014wide,hong2019high,ding2014fully}, bi-layer gratings~\cite{michaels2018inverse}, as well as advanced designs~\cite{zaoui2012cost,michaels2018inverse,sacher2014wide,hong2019high,ding2014fully}. Advanced optimization algorithms~\cite{su2018fully,mak2018silicon}, including inverse design~\cite{michaels2018inverse,hooten2021inverse,sapra2019inverse,son2018high}, are applied to maximize the coupling efficiency.

However, the integration of additional functionality in the coupling device is equally important. This allows for a reduction in the number of optical elements used in the scheme and simplifies it. For grating couplers, significant efforts have been made to enable linear polarization splitting~\cite{taillaert2003compact,streshinsky2013compact,tang2009proposal,zaoui2013cmos,mekis2010grating,watanabe20192,wang2010experimental}, separation by angular orbital momentum~\cite{zhou2019ultra,zhao2019compound,nadovich2016forked,wang2022chip,fatkhiev2021recent}, efficient routing of TE and TM modes in different directions~\cite{zaoui2013cmos,wang2010experimental}, enhancement and collimation of spontaneous emission~\cite{aouani2011plasmonic,steele2006resonant}, focusing of the outcoupled light~\cite{mehta2017precise}, and many other applications~\cite{quaranta2018recent,cheng2020grating,elsawy2020numerical}.

Notably, some of the most advanced and functional  grating coupler designs are widely applied for excitation and manipulation of the surface plasmons. For example, there have been numerous studies on the photonic spin-orbit coupling effect in slit-based gratings for directional coupling of surface plasmons~\cite{lin2013polarization,bliokh2015spin,shitrit2013spin,mueller2014polarization, ding2018bifunctional, huang2013helicity, miroshnichenko2013polarization, yang2020deuterogenic, xu2017polarization,zhang2018polarization,consales2020metasurface,Pham2018,revah2019}, angular orbital momentum splitting~\cite{zhao2019compound}, guided mode wavefront manipulation (near-field focusing)~\cite{chen2020polarization,zhao2019polarization,lee2015plasmonic}, and polarimetry~\cite{zhang2018polarization,lee2018ultracompact}. 
The mentioned effect of photonic spin-orbit coupling that is also utilized in our study is actively employed not only for surface plasmons routing, but for light manipulation in many other optical structures as well. In particular, selectivity by the light handedness was demonstrated for scattering by individual nanoparticles~\cite{o2014spin}, directional emission in metamaterials~\cite{kapitanova2014photonic}, photonic crystal modes routing~\cite{parappurath2020direct}, directional Raman scattering~\cite{guo2019routing}, handedness-controlled waveguide splitters~\cite{espinosa2017chip, zhang2017chip}. Similar structures and ideas are implemented for realization of chiral mirrors~\cite{plum2015chiral}, metasurfaces~\cite{gorkunov2020metasurfaces,Dyakov2020} and resonators~\cite{dyakov2023chiral,baranov2023toward,voronin2022single,kim2023chiral} that also interact differently with light depending on its handedness.

\begin{figure*}
    \centering
    \includegraphics[width=0.65\linewidth]{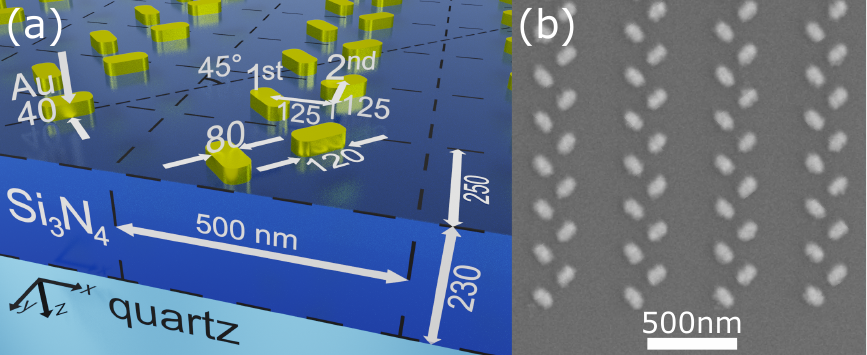}
    \caption{ (a) Schematic of the plasmonic lattice on the surface of a $\mathrm{Si}_3\mathrm{N}_4$ waveguide, deposited on a quartz substrate. The illustrated structure of the unit cell is designed to enable directional excitation of waveguide modes by circularly polarized light. (b) Scanning electron microscope image of the lattice, showing the shape of the fabricated nanoparticles.}
    \label{fig:fig1}
\end{figure*}

Although many of the advanced approaches were originally developed for surface plasmons, their adaptation for photonic waveguides is no less promising. Photonic guided modes that appear similar to surface plasmons, in some applications offer significant advantages. In addition to the absence of Joule losses, dielectric waveguides support not only a single guided mode but multiple guided modes of TE and TM polarizations, opening up wider opportunities in on-chip signal multiplexing/demultiplexing.
In our previous studies~\cite{fradkin2020nanoparticle, fradkin2022plasmonic}, we have shown that the slit-based grating originally designed to rout surface plasmons~\cite{lin2013polarization} can be adapted for manipulation of waveguide modes in dielectric slabs. The unit cell of the adapted grating on top of the waveguide had the similar structure to the original one~\cite{lin2013polarization}, but consisted of elongated plasmonic nanobars instead of the slits.
Such choice was determined by the ability of plasmonic nanoparticles to confine light at deep subwavelength scales, which makes it possible to achieve a strong dipolar response even via few dozen nanometers particles. Such "lumped" elements are extremely convenient for engineering optical structures and metasurfaces with desired properties. For this reason, plasmonic nanoparticle lattices have already been used in various applications, including obtaining hybrid lattice resonances~\cite{Linden2001,jiang2018metasurface,christ2003waveguide,tikhodeev2005waveguide,guan2020engineering,murai2013hybrid}, lasing~\cite{zyablovsky2017optimum,zhou2013,wang2017,schokker2016lasing,schokker2017systematic}, holograms~\cite{zheng2015metasurface,ye2016spin,wei2017broadband,ignatov2018excitation}, flat optics~\cite{yu2014flat,chen2016review,kildishev2013planar}, and many other applications~\cite{Rajeeva2018,kravets2018plasmonic,utyushev2021collective,baburin2019silver}.

In~\cite{fradkin2020nanoparticle,fradkin2022plasmonic} we proposed a lattice with proper positions and orientations of plasmonic nanobars for scattering out waveguide modes into circularly polarized far-field light, which polarization handedness is determined by the direction of mode propagation. The fabricated on-chip light source exhibited 80\% degree of circular polarization of outcoupled TE mode initially emitted by the quantum dots embedded within GaAs waveguide~\cite{fradkin2022plasmonic}. However, the structure we demonstrated had significant limitations. For example, the self-assembled layer of InAs quantum dots required low temperatures to operate and could only excite TE modes, but not TM ones.
The high refractive index of GaAs appeared to promote the excitation of highly confined and lossy edge plasmons that are extremely shape-sensitive and generate high-order multipole moments, which significantly complicates the structure's description. Taken together, these factors indicate the high potential for further improvement of the grating's functionality and performance.

In this study, we further develop functional, polarization-controlled grating coupler based on a lattice of plasmonic nanoparticle. We consider the grating on the surface of a dielectric silicon nitride, $\mathrm{Si}_3\mathrm{N}_3$, slab waveguide, which is one of the widespread platforms for integrated photonics. At room temperature we demonstrate that the structure can couple circularly-polarized light with waveguide modes of both TE and TM polarizations. Furthermore, TE and TM modes coupled with light of specific circular polarizations propagate in opposite directions, enabling both polarization-dependent routing of waveguide modes and potential integration of a light sources with tunable polarization. Specifically, we demonstrate the incoupling of a circularly-polarized beam with up to 95\% routing efficiency for normally incident light, and the outcoupling of waveguide modes with a 97\% degree of circular polarization for normal light propagation. These unique effects provide a convenient grating coupler-based platform for on-chip polarization-division multiplexing and demultiplexing.

\section{Results and Discussion}

\begin{figure*}
    \centering
    \includegraphics[width=\linewidth]{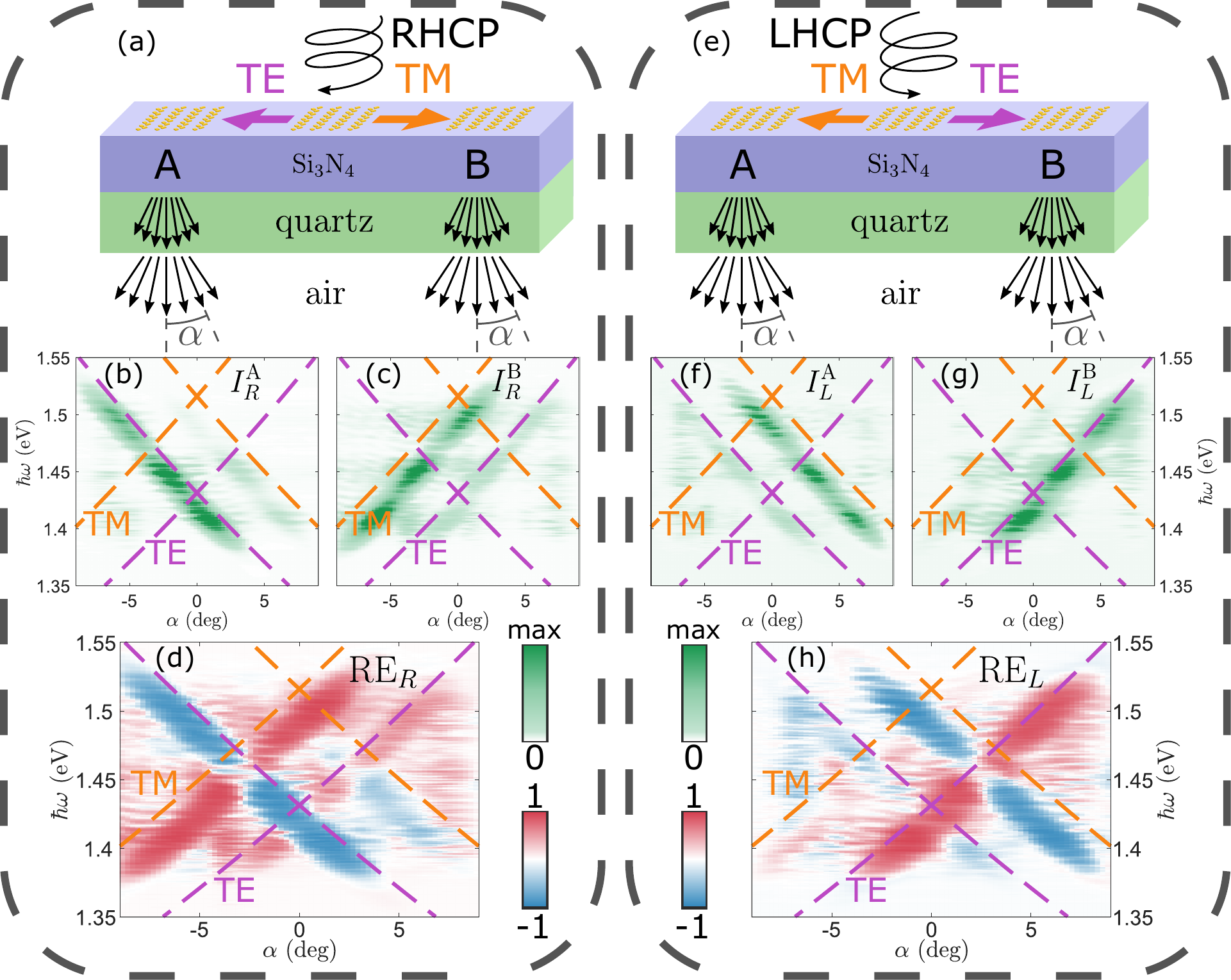}
    \caption{ 
        The structure of three identical $30\times30$~$\mu\mathrm{m}^2$ plasmonic gratings on a $100\mu$m center-to-center distance is analyzed to study incoupling of right-(a-d) and left-(e-h) handed circularly polarized light by the central grating. Incident beam excites TE and TM waveguide modes propagating in opposite directions, which are subsequently outcoupled by the neighboring gratings A and B and detected from beneath the structure. 
        Spectral maps of the outcoupling radiation from the left-sided grating~A~(b,f) and right-sided grating B~(c,g) exhibit distinct lines of waveguide modes. Panels~(d,h) show regularized routing efficiencies,
        reaching up to 95\% for normal incidence. Dashed violet and orange lines represent the dispersion of TE and TM modes in the empty lattice approximation.
    }
    \label{fig:fig2}
\end{figure*}

We work with a lattice of gold nanoparticles on the surface of a silicon nitride, $\mathrm{Si}_3\mathrm{N}_4$ ($\varepsilon_{\mathrm{Si}_3\mathrm{N}_4}=1.99^2$), optical slab waveguide (see Fig.~\ref{fig:fig1}) that was fabricated on a 170~$\mu$m high-purity quartz, $\mathrm{SiO}_2$, substrate ($\varepsilon_{\mathrm{quartz}}=1.55^2$). The thickness of the waveguide was estimated to be approximately $H=230$~nm (for more details, see section~\ref{app:app_wg_modes}).
Unit cell of the structure consists of two perpendicular nanobars $120\times80\times40$~nm${}^3$ aligned at $\pm45\degree$. These sizes allow to bring the longitudinal plasmon resonance closer to the spectrum range of interest ($1.35-1.55$~eV) and ensure the dominance of longitudinal polarizability over the transverse component. Numerical calculations for the polarizability components of the fabricated nanoparticles (see section~\ref{sec:polarizability}) indicate that, in practice, nanobars can be considered as uniaxial scatterers. To couple the lattice with $x$-propagating TE and TM modes, we selected an appropriate $x$-axis period, $p_x=p=500$~nm. Simultaneously, the $y$-axis period was halved to $p_y=p/2=250$~nm to prevent the excitation of $y$-propagating modes. A quarter-period shift along the $x$-axis ($p/4=125$~nm) combined with the proper orientation of the particles allows achieving selective excitation of $x$-propagating waves based on the incident beam's handedness. A similar shift along the $y$-axis is necessary to suppress any parasitic interactions between the two sublattices~\cite{baur2018}.

To investigate the routing properties of the structure, we positioned three identical gratings, $30\times30\mu\mathrm{m}^2$ size, at a center-to-center distance of $100\mu\mathrm{m}$ from each other (see Fig.~\ref{fig:fig2}~(a,e)). To cover the entire spectral region of interest, we illuminated the structure with a circularly-polarized tunable TiSp laser (for the overall excitation spectrum, see section~\ref{app:app_a}). The laser was focused to a spot size of $\approx30\mu$m, covering the central grating. Upon resonant excitation, TM and TE waveguide modes propagated in both directions until they were outcoupled by the neighboring lattices. Spectra of the beams outcoupled to the substrate were in turn measured as a function of the outcoupling angle in the air $\alpha$, $k_x=\frac{2\pi}{\lambda}\sin \alpha$, $k_y=0$ (for details on the experimental setup, see section~\ref{app:app_b}).

Our first objective is to understand how circularly-polarized light incident on a central grating excites waveguide modes of both polarizations. To determine the intensity of the left- and right-propagating modes, we measure the intensity of light that is outcoupled from the left-sided grating A, denoted as $I^\mathrm{A}_\mathrm{R/L}$, and from the right-sided grating B, denoted as $I^\mathrm{B}_\mathrm{R/L}$. The subscripts R and L refer to the incident beam of right-handed circularly-polarization, RHCP, (see Fig.~\ref{fig:fig2} schematic (a) and spectra (b-d) below) and left-handed circularly-polarization, LHCP, (see Fig.~\ref{fig:fig2} schematic (e) and spectra (f-h) below), respectively. In this paper, we use the handedness definition described in section~\ref{app:circular_polarization}. The polarization of the outcoupled light is not analyzed at this stage. To simplify the identification of the observed modes, we demonstrate the so-called empty lattice approximation for the dispersion of TE and TM modes with dashed violet and orange lines, respectively. These lines represent the dispersion of waveguide modes in a pure Si$_3$N$_4$ waveguide without any grating on top, which are formally folded in the first Brillouin zone. The guided modes of the actual structure are slightly red-shifted mainly due to hybridization with localized plasmons of individual plasmonic nanoparticles (see section~\ref{app:app_wg_modes} for details).

The observed spectra clearly demonstrate that RHCP light beam efficiently excites only the TE mode propagating to the left direction (see Fig.~\ref{fig:fig2}~(b)), while it excites the TM mode propagating to the right direction (see Fig.~\ref{fig:fig2}~(c)). Vice versa effect is observed for LHCP illumination, when the TE and TM modes are propagate to the right and to the left, respectively (see Fig.~\ref{fig:fig2}~(f-g)).

Based on the intensities of the outcoupled light, the routing efficiency (RE) of the grating can be calculated for RHCP and LHCP incident light, $\mathrm{RE}_{\mathrm{R/L}}=\frac{I^{\mathrm{B}}_{\mathrm{R/L}}-I^{\mathrm{A}}_{\mathrm{R/L}}}{I^{\mathrm{B}}_{\mathrm{R/L}}+I^{\mathrm{A}}_{\mathrm{R/L}}}$. To avoid noise-related issues arising from near-zero denominator values, we plot the regularized RE, which is calculated based on the intensities with an additional artificial background of $I_\mathrm{bg}=0.03 I_{\mathrm{max}}$ (see~Fig.~\ref{fig:fig2}~(d,h)). It can be noticed that this background does not affect the numerator, but introduces an additional term of $2I_\mathrm{bg}$ to the denominator.
Negative values of routing efficiency (blue colored lines) correspond to left-propagating modes, while positive values (red colored lines) correspond to right-propagating modes, simplifying the understanding of the routing behavior of the plasmonic grating. Specifically, for normal incidence of light, we observe routing efficiency with an absolute value, $|\mathrm{RE}|$, of up to approximately 95\%. The demonstrated routing effect could potentially be used for demultiplexing polarization-encoded signals, polarization state measurement, and other purposes.

The routing effect described above might be qualitatively explained by a simple model. The considered plasmonic nanoparticles are small enough to describe them in electric dipole approximation. In addition, they are designed to serve as uniaxial scatterers in appropriate frequency range ($\approx1.35-1.55$~eV), which allows to describe the in-plane polarizability of the first, $\hat{\alpha}_1$, and second, $\hat{\alpha}_2$, particles (see Fig~\ref{fig:fig1} for the numbering) consisting the unit cell by the following tensors:

\begin{equation}
    \hat{\alpha}_1 \approx 
    \frac{\alpha}{2}
    \begin{bmatrix}
    1 &1\\
    1 & 1
    \end{bmatrix}, \quad
    \hat{\alpha}_2 \approx 
    \frac{\alpha}{2}
    \begin{bmatrix}
    1 & -1\\
    -1 & 1
    \end{bmatrix},
\end{equation}
where $\alpha = \alpha_{ll}$ is the dominating, longitudinal polarizability of the nanobar.

Assuming that dipole moments of the nanoparticles are determined only by the external field of the incident wave we easily derive the following expressions:

\begin{multline}
    \mathbf{P}_1^\mathrm{R/L} = \hat{\alpha}_1 \mathbf{E}^\mathrm{R/L}_{\uparrow z} = 
    \frac{\alpha}{2}
    \begin{pmatrix}
    1 & 1\\
    1 & 1
    \end{pmatrix}\frac{1}{\sqrt{2}}\begin{pmatrix}
    1 \\
    \mp i 
    \end{pmatrix} E^\mathrm{R/L} =
    \\ \frac{\alpha}{2}e^{\mp i \pi/4}\begin{pmatrix}
    1 \\
    1 
    \end{pmatrix}E^\mathrm{R/L},\label{eq:P1}
\end{multline}

\begin{equation}
    \mathbf{P}_2^\mathrm{R/L} = \hat{\alpha}_2 \mathbf{E}^\mathrm{R/L}_{\uparrow z}  = 
  \frac{\alpha}{2}e^{\pm i \pi/4}\begin{pmatrix}
    1 \\
    -1 
    \end{pmatrix}E^\mathrm{R/L},
    \label{eq:P2}
\end{equation}
where $\mathbf{E}^\mathrm{R/L}$ are the fields of right- and left-handed circularly polarized waves, whereas $\mathbf{P}_1$ and $\mathbf{P}_2$ are corresponding dipole moments of the nanoparticles. For the sake of simplicity, we assumed the normal incidence, but it makes a good approximation for the considered range of angles ($\pm 9^\degree$).

\begin{figure*}
    \centering
    \includegraphics[width=\linewidth]{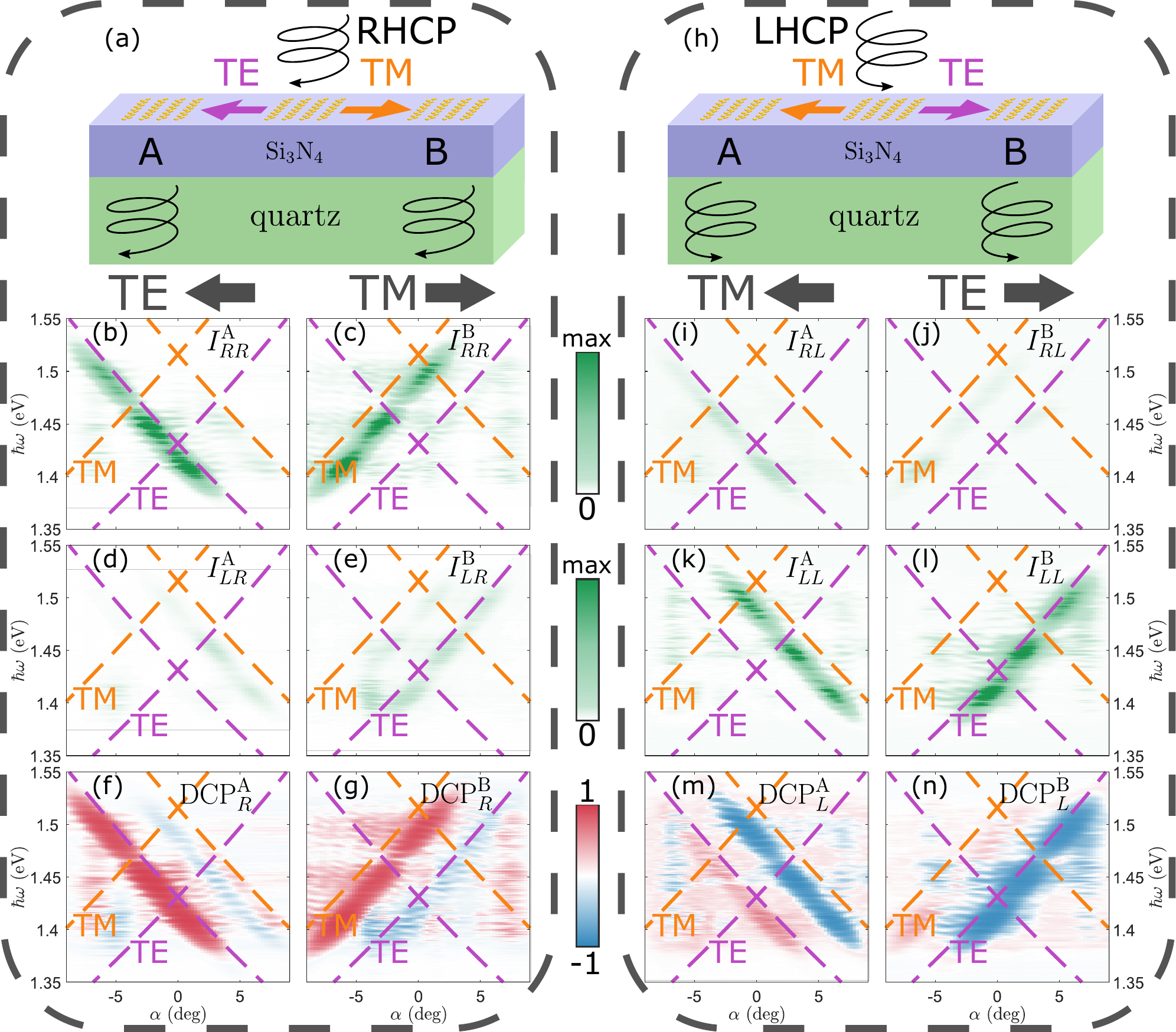}
    \caption{
        The structure of three identical $30\times30\mu\mathrm{m}^2$ plasmonic gratings on a $100\mu$m center-to-center distance is analyzed to study outcoupling of waveguide modes by the lateral gratings~A~and~B. Excitation of waveguide modes by central grating via right-(a-g) and left-(h-n) handed circularly polarized light allows us to measure outcoupling of each mode separately.
        Four columns of spectral maps (b-g) and (i-n) demonstrate the measurements for the outcoupling of left- and right- propagating TE and TM modes.
        In particular, the first row of spectra (b-c,i-j) shows the intensity of RHCP outcoupled light, the second row (d-e,k-l) shows LHCP one, and the third one (f-g,m-n) shows the resulting regularized degree of circular polarization reaching 97\% for normal direction. Dashed violet and orange lines indicate the dispersion of TE and TM modes in the empty lattice approximation.
    }
    \label{fig:fig3}
\end{figure*}

Although plasmonic nanoparticles possess strong, resonant optical response they are weakly coupled with the modes of dielectric slab.
Therefore, plasmonic grating might be considered just as an interface that allows either exciting or scattering out waveguide modes, but not significantly changing their dispersion, polarization or field distribution. In this scope, we are able to assess the amplitudes of waveguide modes excited by the primarily induced dipole moments from Eq.~\ref{eq:P1}-\ref{eq:P2}. In order to do that, we take the field of the mode dual to the considered one (propagating in opposite direction in our case) at the positions of the nanoparticles ($x_{1/2}=\mp p/8$) and calculate its product with the dipole moments~\cite{vainshtein1988}. Since we consider the eigenmodes of the hybrid structure in the vicinity of the $\Gamma$ point it is possible to approximate the wavevectors of both waveguide modes by the Bragg vector $\beta_\mathrm{TE}\approx \beta_\mathrm{TM}\approx 2\pi/p$. Let us estimate the amplitude of the left-propagating TE and TM modes:

\begin{multline}
    A_\mathrm{TE}^\leftarrow(\mathrm{R/L})\propto \sum_{n={1,2}}\mathbf{E}_{\mathrm{TE}}^\rightarrow e^{i\beta_\mathrm{TE}x_n} \cdot\mathbf{P}_n^\mathrm{R/L}=\\
    \begin{pmatrix}
        0 \\ 1
    \end{pmatrix}e^{i\frac{2\pi}{p}\frac{-p}{8}}
    \cdot\mathbf{P}_1^\mathrm{R/L}+\begin{pmatrix}
        0\\ 1
    \end{pmatrix}e^{i\frac{2\pi}{p}\frac{p}{8}}
    \cdot\mathbf{P}_2^\mathrm{R/L}=\\
    \left\{\begin{matrix}
        -i\alpha E^{\mathrm{R}}&\ \mathrm{for \ RHCP}\\
        0& \ \mathrm{for \ LHCP}
    \end{matrix}\right.,
\end{multline}

\begin{multline}
    A_\mathrm{TM}^\leftarrow(\mathrm{R/L})\propto \sum_{n={1,2}}\mathbf{E}_{\mathrm{TM}}^\rightarrow e^{i\beta_\mathrm{TE}x_n} \cdot\mathbf{P}_n^\mathrm{R/L}=\\
    \begin{pmatrix}
        1 \\ 0
    \end{pmatrix}e^{i\frac{2\pi}{p}\frac{-p}{8}}
    \cdot\mathbf{P}_1^\mathrm{R/L}+\begin{pmatrix}
        1\\ 0
    \end{pmatrix}e^{i\frac{2\pi}{p}\frac{p}{8}}
    \cdot\mathbf{P}_2^\mathrm{R/L}=\\
    \left\{\begin{matrix}
        0&\ \mathrm{for \ RHCP}\\
        \alpha E^{\mathrm{L}} & \ \mathrm{for \ LHCP}
    \end{matrix}\right..
\end{multline}

As we can see, an approximate assessment confirms experimentally demonstrated fact that left-propagating TE mode is excited only by RHCP light,  and TM mode only by LHCP one. In a similar way it might be shown that right-propagating modes might be excited only by the light of different handedness.

In our derivations we have neglected interactions of sublattices and optical feedback of the waveguide modes. Nevertheless, appropriate considerations for the lattice of the observed type does not qualitatively affect the routing effect~\cite{fradkin2020nanoparticle, baur2018}, but allow to quantitatively assess the amplitude of waveguide excitation and shift of the resonant line.

Summing up the estimations, circularly polarized light induces dipole moments with $\pm\pi/2$ phase shift between the particles depending on the handedness and considered component ($x$ or $y$) of dipole moment. Concurrently, $\approx\pm\pi/2$ phase delay is gained by the waveguide modes propagating the $p/4$ distance between the particles. As a result, some modes are efficiently excited due to constructive interference ($\pi/2-\pi/2=0$), whereas some of them are suppressed due to destructive one ($\pi/2+\pi/2=\pi$).

We have investigated the behavior of the grating for light incoupling, and now we will focus on the outcoupled light and its polarization (see Fig.~\ref{fig:fig3}). At each grating (A and B), we separate the outcoupled signal into LHCP and RHCP contributions. For example, the total intensity of the A-grating outcoupled waveguide modes (originally excited by RHCP), denoted as $I^\mathrm{A}_\mathrm{R}$ (see Fig.~\ref{fig:fig2}~(b)), can be divided into RHCP and LHCP contributions, represented by $I^\mathrm{A}_\mathrm{RR}$ and $I^\mathrm{A}_\mathrm{LR}$ respectively, as shown in Fig.~\ref{fig:fig3}~(b,d). It can be observed that the left-propagating TE mode, initially excited by RHCP, is predominantly outcoupled in the substrate with RHCP as well, with $I^\mathrm{A}_\mathrm{RR}\gg I^\mathrm{A}_\mathrm{LR}$. This phenomenon applies to both circular polarizations of incident light and both waveguide modes as well (see Fig.~\ref{fig:fig3}~(a,h) for a schematic of the four possible combinations).
From the analytical estimations above, it can be understood that the coupling of waveguide modes with a beam is fully determined by the direction of local in-plane field rotation. Therefore, it is irrelevant whether the beam is incident on the structure or outcoupled from it. Since the excitation beam and the outcoupled beam interact with identical gratings through the same mode and propagate in the same direction (downwards), they possess the same handedness of polarization.
In Fig.~\ref{fig:fig3}~(d), a suppressed line of left-propagating TM mode can be observed. All the other measured spectra (panels (c,e,i-l)) demonstrate a similar effect, where the dominant mode maintains its handedness while the suppressed mode undergoes a change in polarization.

This effect can be conveniently illustrated by the degree of circular polarization, $\mathrm{DCP}^{\mathrm{A/B}}_\mathrm{R/L}=\frac{I^{\mathrm{A/B}}_{\mathrm{RR/RL}}-I^{\mathrm{A/B}}_\mathrm{LR/LL}}{I^{\mathrm{A/B}}_\mathrm{RR/RL}+I^{\mathrm{A/B}}_\mathrm{LR/LL}}$, of the outcoupled light from each grating, A/B, for each polarization of illumination, R/L. Similar to the routing efficiency, we regularize the plotted DCP by adding a small value of background intensity $I_\mathrm{bg}=0.03 I_{\mathrm{max}}$ in the calculations (see Fig.~\ref{fig:fig3}~(f-g,m-n)). The degree of circular polarization for the modes with dominant intensity reaches up to 97\% for normal propagation.

The behavior of the plasmonic grating demonstrated in this study enhances our ability to manipulate and control light on a chip. This makes it a promising interface for coupling circularly-polarized light with waveguide modes. Unlike TM-polarized surface plasmons, which only operate with modes of one polarization, our grating allows for operation with modes of both polarizations and potentially even with high order waveguide modes. Waveguide modes of dielectric slabs do not suffer from Joule losses, making them preferable for use in optical communications on a chip. The use of plasmonic nanoparticles should not be a barrier in terms of energy efficiency, as dielectric grating couplers also have significant losses on scattering and backreflection.

\section{Conclusion}

In this work, we have shown that our proposed plasmonic grating coupler is suitable for both incoupling and outcoupling of waveguide modes. In the first case, it allows for exciting the mode propagating in a single direction by normally incident light and enables circular-polarization controlled routing of the modes propagation direction with 95\% efficiency. This paves the way for demultiplexing incoming signals based on their handedness. In the second case, the plasmonic grating allows for outcoupling light from the chip with up to 97\% degree of circular polarization for normal propagation. The handedness of the outcoupled light can be controlled by the polarization and propagation direction of the waveguide modes. This paves the way for an on-chip light source with arbitrary demanded polarization and multiplexing several channels based on the handedness of the output signal. Overall, the plasmonic grating coupler holds great potential for a wide range of functional devices at the interface of integrated and free-propagating beam optics.

\section{Experimental Section}

\subsection{Spectrum of excitation laser}

\label{app:app_a}

\begin{figure}[h!]
    \centering
    \includegraphics[width=0.9\linewidth]{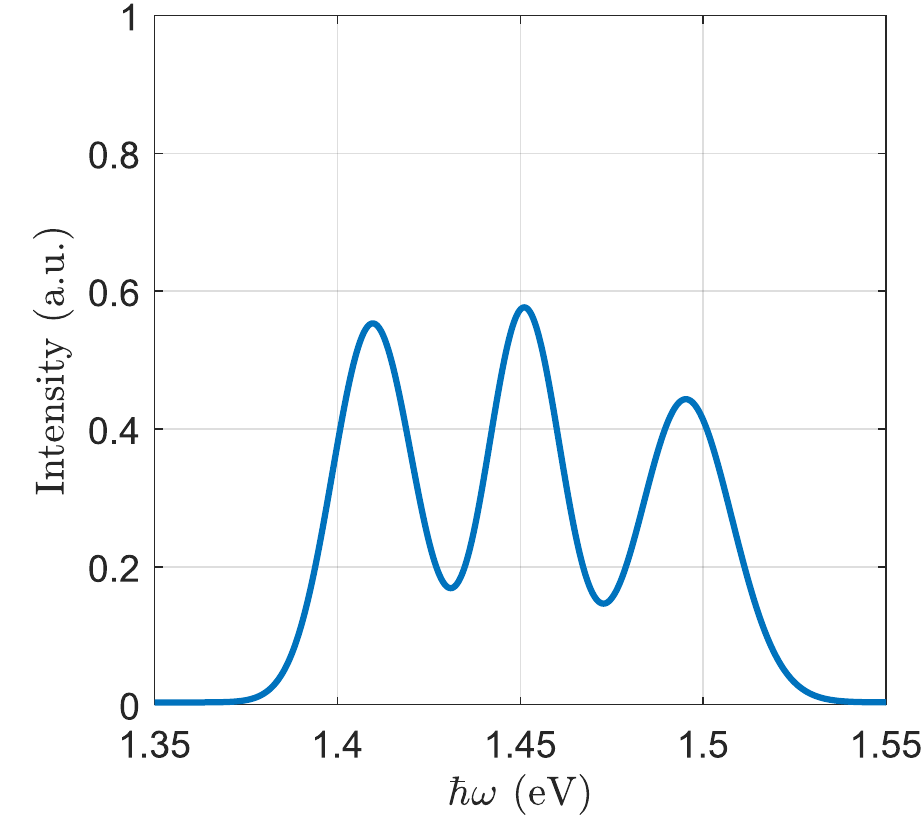}
    \caption{ Estimated cumulative spectrum of a tunable laser.}
    \label{fig:fig_app1}
\end{figure}

In Fig.~\ref{fig:fig_app1} we demonstrate the estimation of the sum of the three wavelength-spaced spectra of a tunable laser that were employed for structure excitation. It helps us to understand the origin of the "horizontal line" in experimentally measured spectral maps. The given graph does not represent any direct measurement, but their approximate fitting.

\subsection{Experimental setup}

\label{app:app_b}

\begin{figure}[h!]
    \centering
    \includegraphics[width=\linewidth]{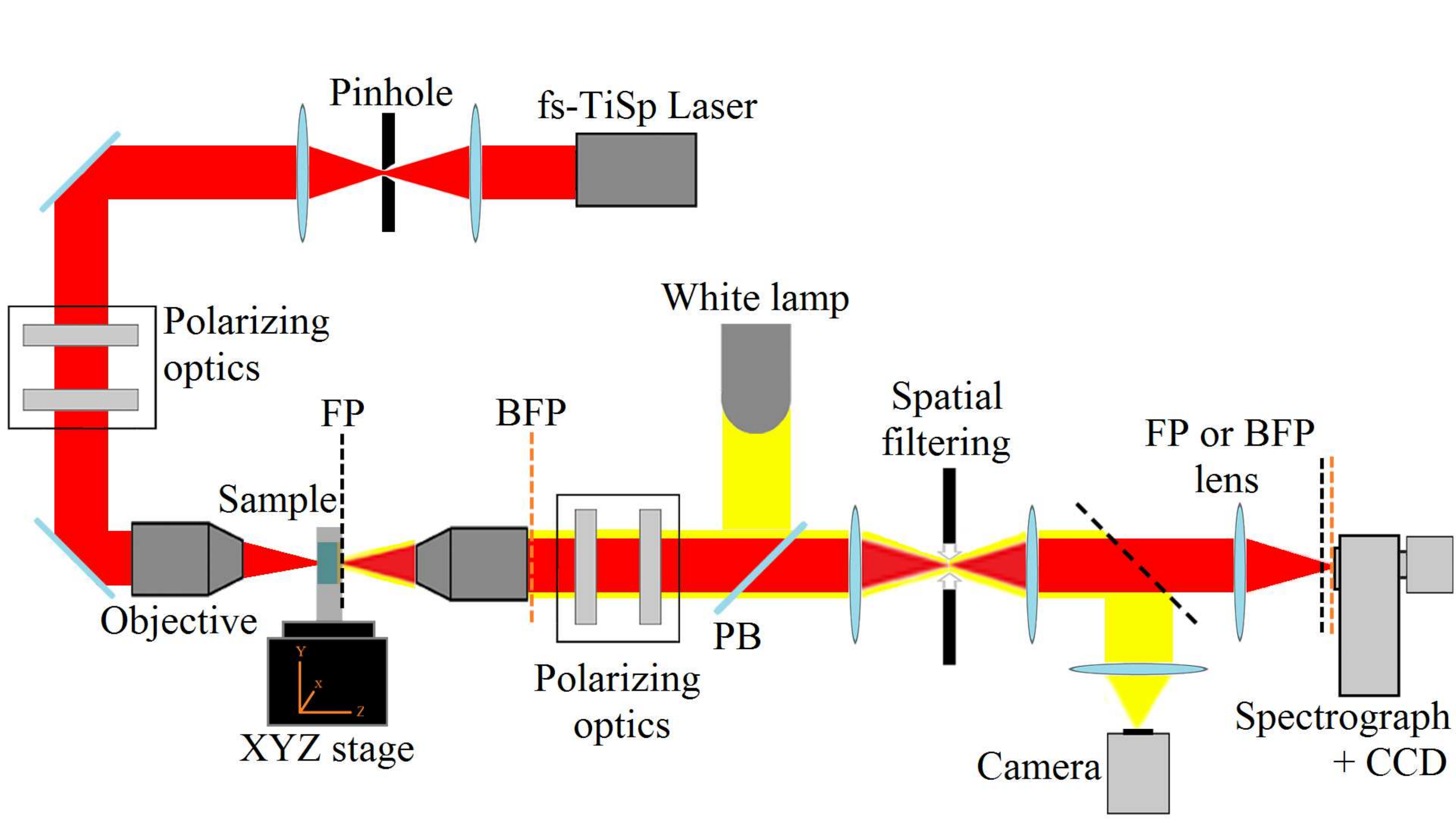}
    \caption{ Scheme of the experimental setup for optical measurements with spatial and angular resolution. When recording images in real space and maps of emission spectra with angular resolution, the focal plane (FP) and back focal plane (BFP) of the Mitutoyo objective were projected by the optical system onto the slit of the image spectrometer and a mirror or a diffraction grating (300~gr/mm) were used, respectively. Spatial filtering of the signal from plasmon gratings was carried out using a double slit. A broadband linear polarizer and an achromatic phase plate~$\lambda/4$ are used to measure the polarization of the radiation. The sample is placed on a precise XYZ stage. A white lamp and an optical camera (The Imaging Source) are used to visualize the plasmon gratings. FP: focal plane; BFP: back focal plane; PB: plate beamsplitter (Thorlabs, 10:90).}
    \label{fig:fig_app2}
\end{figure}

Angularly resolved measurements were conducted using a back focal plane (BFP) imaging setup with resonant excitation by a Spectra-Physics Tsunami tunable femtosecond TiSp laser. In order to obtain dispersion spectra across a wide range of $k$-vectors, the laser wavelength was varied between 810-890~nm. The laser beam was focused onto the plasmonic grating using an optical lens system consisting of a Mitutoyo M Plan APO 10X and a 30 $\mu$m pinhole, resulting in a spot diameter of 30~$\mu$m. The intensity profile of the laser spot had a near top-hat shape, and the excitation density was set at 70 W/cm$^2$. The desired polarization state of the pump laser beam was achieved by combining a linear polarizer from Newport optics with a $\lambda/4$ achromatic plate from Thorlabs. Exciting light reached neighboring plasmonic gratings through TE and TM guided modes is finally got outcoupled by them into a signal, which is recorded in the transmission geometry. Spatial filtering of the signal was performed using an intermediate double slit. The signal was captured using a Princeton Instruments Acton SP-2500i imaging spectrometer coupled to a thermoelectrically cooled imaging CCD camera (Princeton Instruments PIXIS 100) with an exposure time of 30~s. The count rate was about 150~Kcps. To measure the polarization of the signal radiation from the plasmon gratings, a combination of a linear polarizer from Newport optics and a $\lambda/4$ achromatic plate from Thorlabs was used. The sample was positioned on a precise XYZ stage at room temperature. The experimental setup is illustrated in the Fig~\ref{fig:fig_app2}. The optical setup provided spatial, angular, and spectral resolutions of 2~$\mu$m, 0.50$\degree$, and 0.5~meV, respectively, with an error in determining the degree of radiation polarization not exceeding~1.5\%.

\subsection{Atomic force microscopy of the sample}
\label{sec:AFM}

\begin{figure*}
    \centering
    \includegraphics[width=0.9\linewidth]{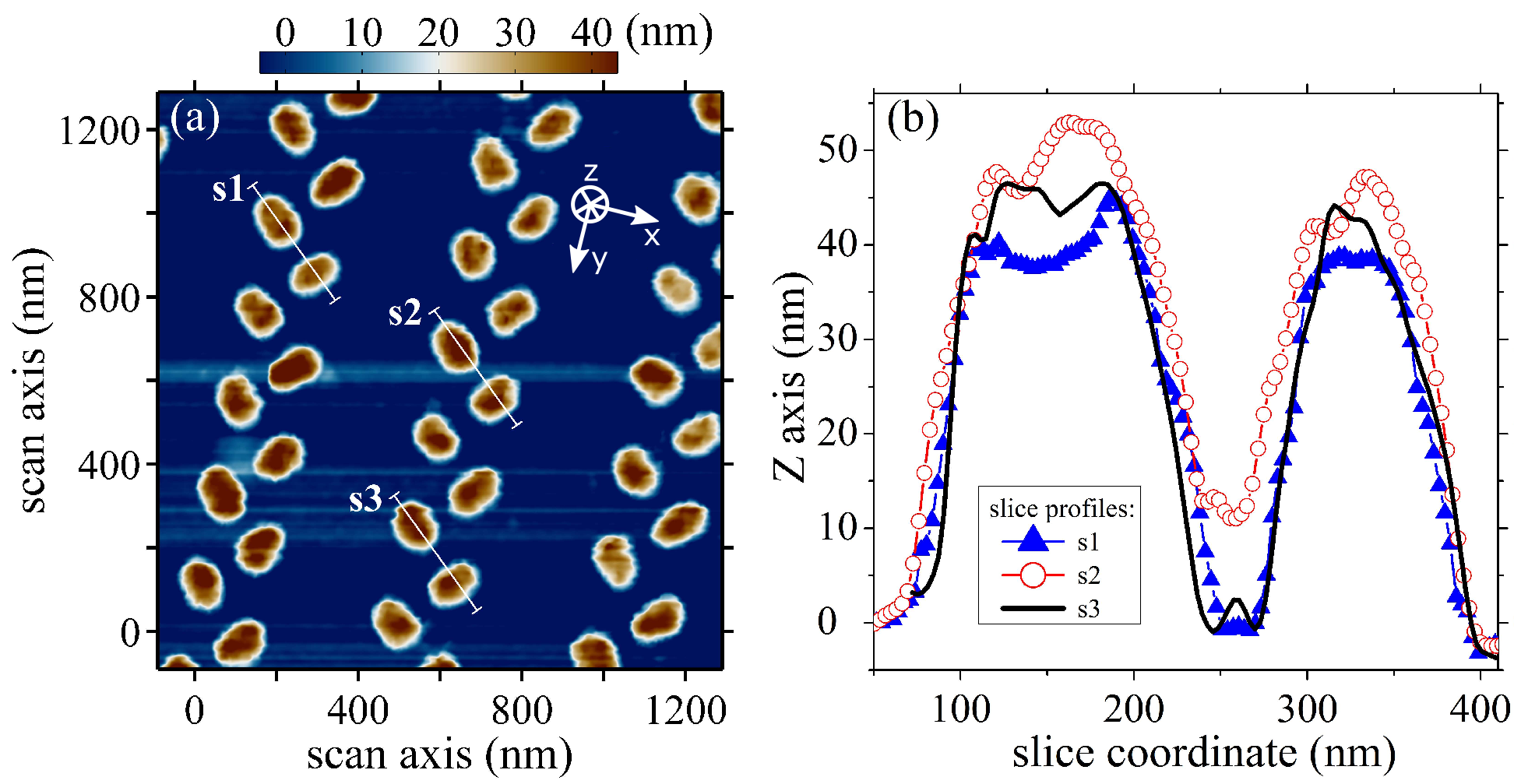}
    \caption{ (a) A fragment with a size of 1.4$\times1.4 \mu\mathrm{m}^2$ of a 2D plasmonic grating measured by AFM; (b) Slices of gold nanoparticles along the main axes of the nanobars, designated in panel (a) as s1, s2, and s3.}
    \label{fig:AFM}
\end{figure*}

The shape and size of the particles we additionally measured using atomic force microscopy (AFM) to confirm the results obtained by SEM. Fig.~\ref{fig:AFM}~(a) shows a 1.4$\times$1.4 $\mu\mathrm{m}^2$ fragment of a two-dimensional (2D) plasmonic lattice. The profile of the gold nanoparticles was obtained  using a whisker-type cantilever with a high aspect ratio ($>1:10$), a cantilever opening angle of $<10\degree$, and a radius of curvature of 10~nm. The scanning step of the AFM microscope was 5~nm. The angle of about $13.5\degree$ between the horizontal scanning axis and the direction of the main $x$ axis is due to the disorientation of the sample on the microscope stage and the direction of its scanning. Fig.~\ref{fig:AFM}~(a) shows that gold nanoparticles have an elongated shape with rather smooth corners. The measured lattice periods along the $x$ and $y$ axes are $p_x \approx 495$~nm and $p_y \approx 230$~nm, respectively, which generally corresponds to both the designed dimensions and SEM measurements. The angles between the nanoparticles in the unit cell are 90$\degree$, and between them and the $x$ and $y$ axes are 45$\degree$, which is important for coupling waveguide modes with the external circularly polarized light field.

Figure~\ref{fig:AFM}~(b) shows three sections of gold nanoparticles that are indicated in Fig.~\ref{fig:AFM}~(a). The measured sections show that the walls of the nanoparticles are inclined to the waveguide plane by $63\degree-68\degree$ degrees. The sizes of nanoparticles measured at their half-height are $135$~nm and $90$~nm, which is also close to dimension extracted from SEM. The distance between their geometrical centers is $173$~nm, which is in a good agreement with the expected value of $125\cdot\sqrt{2}\approx 177$~nm. What is especially important, the demonstrated sections allow to estimate the height of nanoparticles that turns out to be approximately $40$~nm.
Thus, the fabricated plasmonic gratings generally demonstrates good agreement of the actual dimensions characteristics with the initially designed ones.

\section{Methods}

\subsection{On handedness definition}

\label{app:circular_polarization}

In this study, we employ so-called alternative definition of light handedness. We associate right- (left-) handed circular polarization with an electromagnetic wave, which electric field forms right- (left-) handed helix in space. It might be easily shown that under $e^{-i\omega t}$ convention light modes propagating along and against $z$-axis ($\uparrow z$ and $\downarrow z$ correspondingly) right- and left-handed helixes correspond to following field distributions:

\begin{equation}
    \mathbf{E}^{\mathrm{R}}_{\uparrow z} = \frac{1}{\sqrt{2}}\begin{pmatrix}1\\-i \end{pmatrix}e^{i k z}, \quad  
    \mathbf{E}^{\mathrm{R}}_{\downarrow  z} =  \frac{1}{\sqrt{2}}\begin{pmatrix}1\\+i \end{pmatrix}e^{-i k z},
\end{equation}

\begin{equation}        
    \mathbf{E}^{\mathrm{L}}_{\uparrow z} =  \frac{1}{\sqrt{2}}\begin{pmatrix}1\\+i \end{pmatrix}e^{i k z}, \quad
    \mathbf{E}^{\mathrm{L}}_{\downarrow  z} =  \frac{1}{\sqrt{2}}\begin{pmatrix}1\\-i \end{pmatrix}e^{-i k z},
\end{equation}
where vectors consist of fields projections on $x$ and $y$ axes, $k$ is the wavevector in a host medium.

\subsection{Plasmonic nanobar polarizability tensor}

\label{sec:polarizability}

The design of the shape of plasmonic nanoparticles aims to achieve a structure close to a uniaxial scatterer. To do it by tuning the longitudinal plasmonic resonance to an energy of $\approx 1.45$~eV it is necessary to fabricate particles of approximately $100\times45\times25$~nm$^{3}$. However, in practice, the actual size of plasmonic nanobars may deviate significantly from the desired dimensions.

To accurately assess the mean length and width of the nanobars, we averaged measurements obtained from scanning electron microscopy, SEM,  (see Fig.~\ref{fig:fig1}~(b)) and atomic force microscopy, AFM, (see~sec.~\ref{sec:AFM}). Both methods provide close estimations (approximately $120\times80$~nm${}^2$) for the in-plane sizes of nanobars, which is slightly larger than originally planned dimensions. Regarding the thickness of the gold layer, we could only extract this information from AFM measurements, which indicated a height of approximately $40$~nm, significantly deviating from the expected value of $25$~nm.

\begin{figure*}
    \centering
    \includegraphics[width=0.9\linewidth]{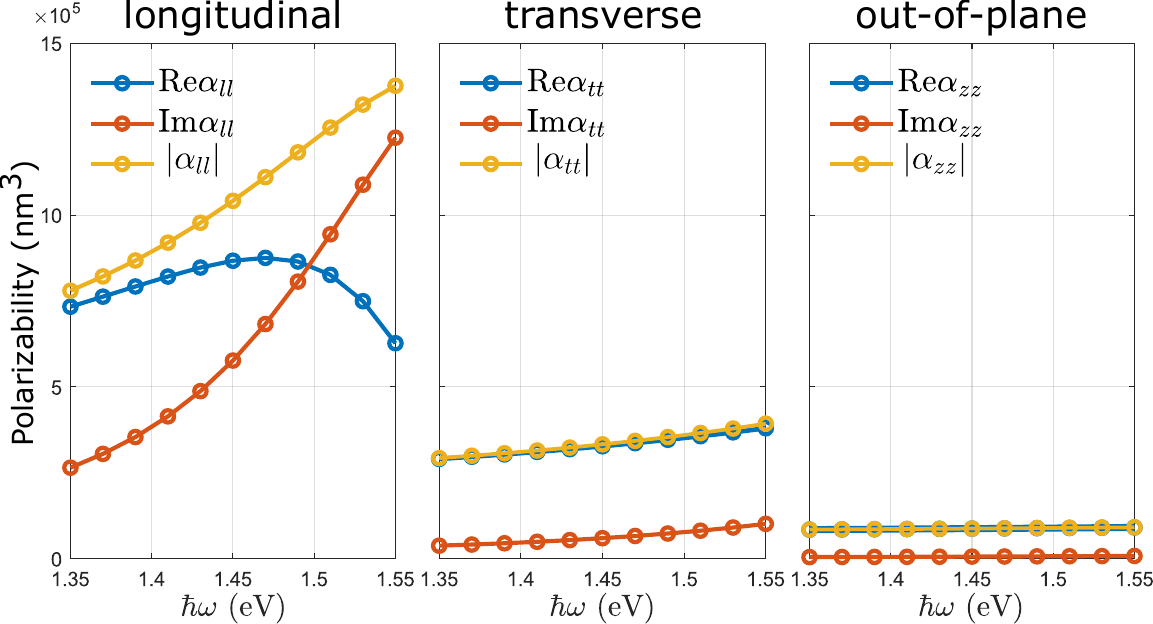}
    \caption{ Longitudinal, transverse, and out-of-plane components of $120\times80\times40$~nm${}^3$ nanobar polarizability tensor. The predominance of longitudinal polarizability over transverse polarizability is clearly observed.}
    \label{fig:fig_polarizability}
\end{figure*}

Based on the measured sizes, we calculated the polarizability tensor of the fabricated nanoparticles to assess their optical properties. This was done numerically via the current-injection-based technique, that allows for the consideration of nanoparticles of any shape in any environment~\cite{fradkin2019fourier}. In our case, the golden nanoparticles were considered in air on the surface of a Si${}_3$N${}_4$ layer. The gold permittivity was described by Johnson\&Christy optical constants~\cite{JohnsonChristy1972}, while the permittivity of the underlying, Si${}_3$N${}_4$, layer was taken as $\varepsilon_{\mathrm{Si}_3\mathrm{N}_3}=(1.99+0.0028i)^2\approx3.96 + 0.02i$. In-plane cross section of the stadium-shaped nanoparticle we composed from $40\times80$~nm${}^2$ body rectangle sandwiched between two 80~nm-diameter semicircles.

Spectra of the main components of $120\times80\times40$~nm${}^3$ nanoparticle polarizability tensor are shown in Fig.~\ref{fig:fig_polarizability}. We observe that the longitudinal component of the polarizability tensor dominates two to three times over the transverse component in the spectrum range of interest, confirming that nanobars are suitable for our purposes. Despite the deviation in nanoparticle sizes leading to a blue shift of the resonant line into the visible range, this does not qualitatively alter the ratio of longitudinal to transverse polarizabilities within the studied spectrum range.

\begin{figure}[h!]
    \centering
    \includegraphics[width=\linewidth]{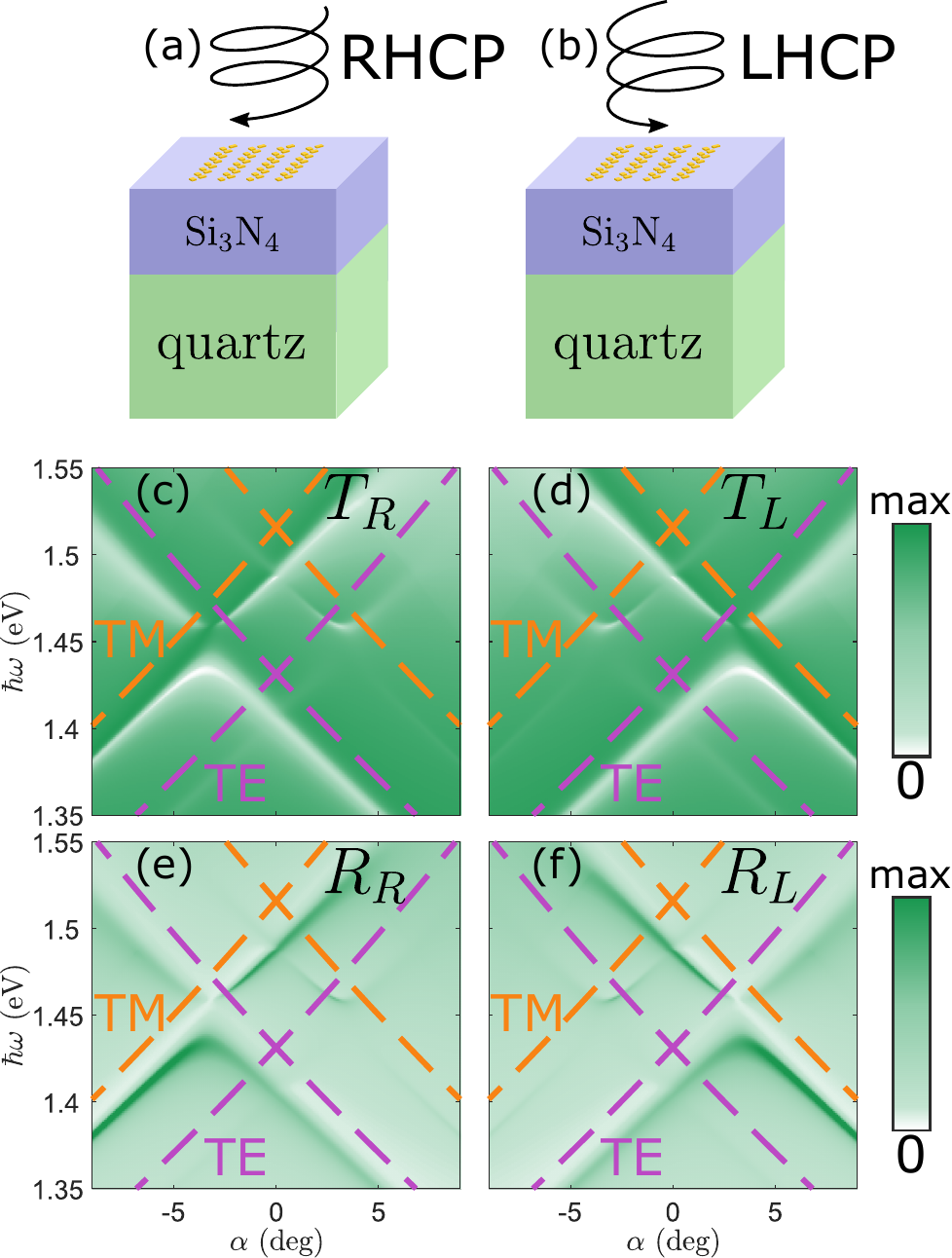}
    \caption{ Schematic of structure considered theoretically under illumination of right-(a) and left-(b) handed circularly polarized light beams and corresponding spectra~(c,e) and~(d,f) below them. Panels (c-d) and (e-f) show transmission and reflection spectra for the main diffraction channel. Hybrid modes correspond to the dips in transmission and peaks in reflection. Dashed violet and orange lines indicate the dispersion of TE and TM modes in the empty lattice approximation.}
    \label{fig:fig_app0}
\end{figure}

\subsection{Numerical spectra of plasmonic gratings}
\label{app:app_wg_modes}

Together with experimental measurements, we have also conducted numerical calculations to analyze the optical properties of the structure. To do this, we utilized a previously developed theoretical approach~\cite{fradkin2019fourier,fradkin2020nanoparticle}, which combines the coupled dipole approximation for plasmonic nanoparticles and rigorous coupled wave analysis for the multilayer structure. This approach enabled us to plot the main diffraction channel transmission and reflection for both right-handed circularly polarized (RHCP) light (see Fig.~\ref{fig:fig_app0}~(a,c,e)) and left-handed circularly polarized (LHCP) light (see Fig.~\ref{fig:fig_app0}~(b,d,f)) incident on an infinite lattice.

From the spectra, it is seen that the excitation of hybrid modes in the structure leads to a decrease in transmission (see Fig.~\ref{fig:fig_app0}~(c-d)) and an increase in reflection (see Fig.~\ref{fig:fig_app0}~(e-f)). The energy of these hybrid modes is mostly lower than that predicted by the empty lattice approximation, due to the hybridization of the original waveguide modes of the dielectric slab with the localized plasmon resonances of nanoparticles. It is important to note that the interaction with the lattice also couples the left-propagating TE mode with the right-propagating TM mode, and vice versa, the right-propagating TE mode with the left-propagating TM mode. In the region of anticrossing, there is a strong coupling between the guided modes, making it impossible to associate the hybrid mode in this area with either TE or TM mode. In the experimentally measured spectra, we do not observe strong coupling effects. One of reasons of this difference is that hybrid modes are eigenmodes of an infinite lattice on the surface of the dielectric slab. In practice, these modes are scattered or reflected at the edge of the finite lattice and continue to propagate beyond the grating as pure waveguide modes of the dielectric slab. When such a mode reaches the edge of a neighboring lattice, it excites hybrid modes again, which subsequently leak out from the structure. The precise analysis of this process is extremely complicated; however, an attempt to provide some numerical estimations can be found, for instance, in~\cite{fradkin2022plasmonic}.

To estimate the true thickness of the Si${}_3$N${}_4$ waveguide, we compared the dispersion of the theoretically calculated hybrid modes with the experimentally measured ones. Such comparison gave us 230~nm thickness of the silicon nitride layer.

\section*{Acknowledgements.} 
The work of I.M.F, S.A.D. and N.A.G. was supported by the Russian Science Foundation (Grant No. 22-12-00351).
The work of A.A.D. and V.D.K. was supported by the Russian Science Foundation (Project
No. 19-72-30003).
The work of A.V.K. was supported by the Ministry of Science and Higher Education of the Russian Federation, STATE TASK No. 075-01304-23-00.
We acknowledge support of  Republic of Bashkortostan grant NOC-RMG-2021.
V.N.A. acknowledges support of Skoltech Next Generation Program, grant number 1-NGP-1064.
I.M.F. acknowledges support by the Foundation for the Advancement of Theoretical Physics and Mathematics “BASIS” on the early stage of this work.

The authors acknowledge S. G. Tikhodeev for fruitful discussions, D. Y. Kazmin for
assistance in experimental work,  and V. V. Ryazanov for valuable contribution. I.M.F. and S.A.D
designed the structures, conducted theoretical computations, and
prepared the manuscript. A.A.D. measured the optical properties of
the structures. 
A.V.K. fabricated the samples. V.D.K., V.N.A and N.A.G. supervised the experimental and theoretical parts of the study correspondingly.

\section*{Keywords}

Chiral light, optical routing, optical waveguide, grating coupler, metasurface, plasmonics, nanophotonics.


%

\end{document}